\documentclass[aps,prb,12pt]{revtex4}
\usepackage{amsfonts,amssymb,amsmath}
\usepackage{graphicx}
\usepackage{yllmath}
\usepackage[usenames,dvipsnames]{color}

\usepackage{subfigure}

\begin{document}
\title{Visualizing Angular Momentum Eigenstates using the Spin Coherent State Representation}
\author{Yen Lee Loh}
\author{Monica Kim}
\affiliation{Department of Physics and Astrophysics, University of North Dakota, Grand Forks, ND  43202, USA}
\date{Original submission 2013-8-20; revised 2013-9-17}
\begin{abstract}
Orbital angular momentum eigenfunctions are readily understood in terms of spherical harmonic wavefunctions.  However, the quantum mechanical phenomenon of spin is often said to be mysterious and hard to visualize, with no classical analogue.  Many textbooks give a heuristic and somewhat unsatisfying picture of a precessing spin vector.  Here we advocate for the ``spin wavefunction'' in the spin coherent state representation as a striking, elegant, and mathematically meaningful visual tool.
We also demonstrate that cartographic projections such as the Hammer projection are useful for visualizing wavefunctions defined on spherical surfaces.
\end{abstract}
\maketitle

Many quantum mechanics textbooks
\cite{griffithsBook,millerBook,liboffBook}
 provide visual representations of orbital angular momentum eigenstates $\ket{lm}$ in terms of their real-space wavefunctions, which are the spherical harmonics $Y_{lm} (\theta, \phi)$.  Spin, however, is said to be a mysterious quantum phenomenon with no classical analogue.  Spin angular momentum eigenstates $\ket{sm}$ are visualized crudely as a semiclassical spin vector of length $\sqrt{s(s+1)} \hbar$ whose tip precesses in a horizontal circle such that $S_z=m\hbar$, as in Fig.~\ref{SpinCartoon}:

	\begin{figure}[!h]
		\centering
		\includegraphics[width=0.4\columnwidth,trim=70mm 90mm 0mm 40mm,clip]{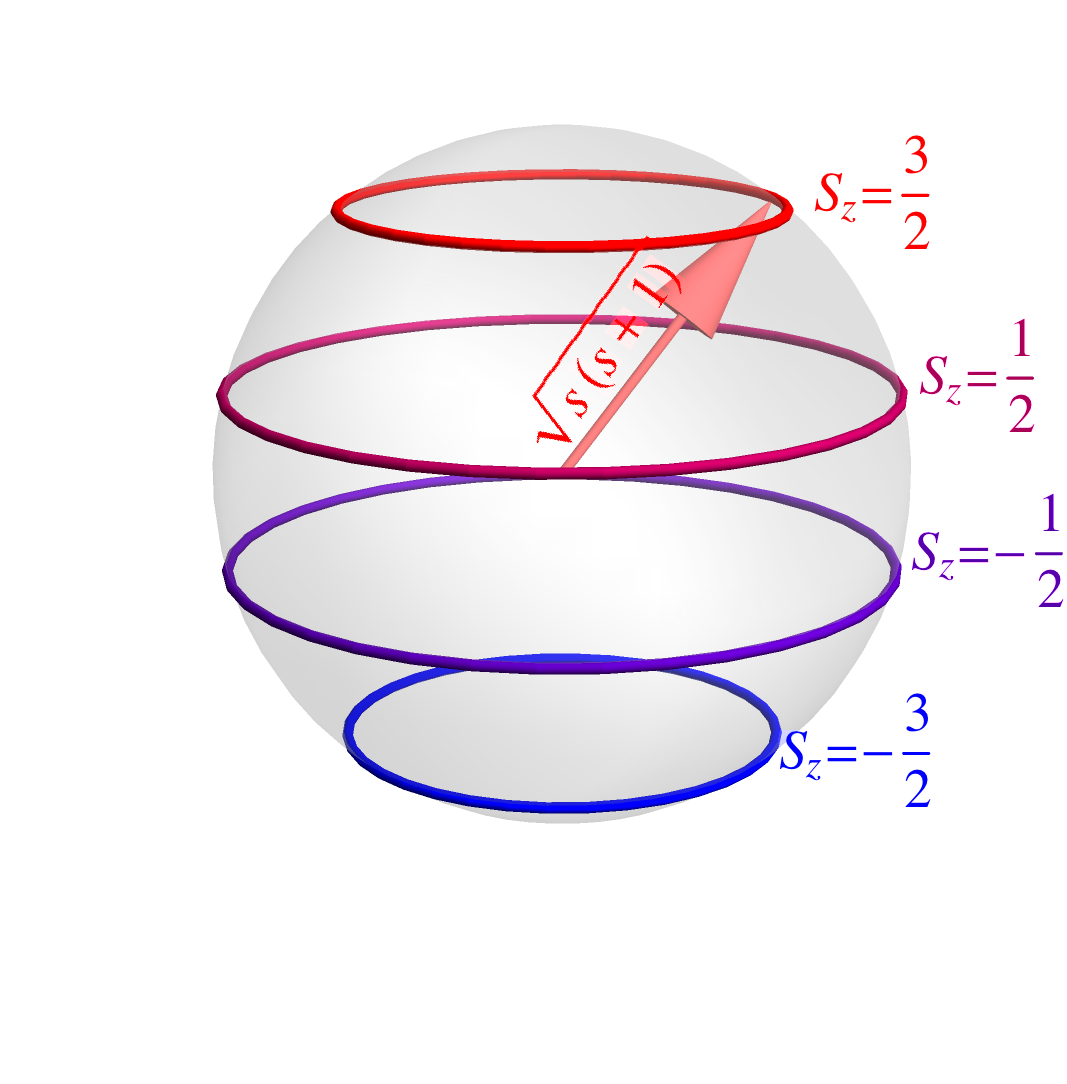}
	  \caption{
	  	Textbook visualization of the spin state $\ket{s=\tfrac{3}{2},m=\tfrac{3}{2}}$.
	  	\label{SpinCartoon}
		}
	\end{figure}

The bulk of the topic is taught in a highly abstract way involving operator algebra and commutation relations.  When confronted with this, students sometimes ask: ``We can do the algebra, but what does it mean?  What does the wavefunction of the spin actually look like?''

Various authors have made attempts to demystify spin by showing how it emerges from Dirac theory\cite{foldy1950,mendlowitz1958,ohanian1986,mita2000,bowman2008,ohanian2009} or Pauli-Schr{\"o}dinger theory\cite{hestenes1979,bowman2008,ohanian2009}, or by making analogies with rotating classical systems\cite{young1976}.  For example, the energy-momentum tensor associated with the Dirac field typically contains a circulating flow of energy corresponding to orbital angular momentum plus spin angular momentum\cite{ohanian1986}.  In principle, one could attempt to gain a visual feel for spin by plotting the Dirac field, which consists of four complex functions in three dimensions.  However, this is restricted to spin-half particles, and the mathematics is beyond the level of most undergraduates.
Here we take a simpler viewpoint, treating spin as a fundamental axiomatic quantity that is unrelated to any position variables, but can still be visualized as a complex function in two dimensions.

In this article we show that any $\ket{sm}$ state can be visualized, in a precise, elegant, and physically meaningful way, as a wavefunction $F_{sm} (\theta, \phi)$ in the spin coherent state representation (SCSR).  For brevity, we will call $F_{sm} (\theta, \phi)$ the ``spin wavefunction'' from now on.  Unlike the $Y_{lm} (\theta, \phi)$, which are meaningful only for integer values of the angular momentum quantum number $l=0,1,2,3,\dotsc$, the $F_{sm} (\theta, \phi)$ are well defined for both integer and half-integer values ($s=0, \tfrac{1}{2}, 1, \tfrac{3}{2}, 2, \dotsc$).  We further show how this picture readily provides visualizations of the sign change of half-integer spin wavefunctions under $2\pi$ rotations, and of physical phenomena such as Larmor precession.

\section{Spin coherent states}
Spin coherent states are well known in the literature \cite{radcliffe1971,arecchi1972,perelomov1972,klauder1979,klauderBook,auerbachBook,aravind1999}, where their primary purpose is for constructing the spin coherent state path integral.  Here we introduce them in a way that is useful for our purposes, and we briefly state their properties.
The reader should not be daunted by the math, which is presented here for completeness but is unnecessary for conceptual understanding.
For convenience we will set $\hbar=1$.

Consider a spin with a fixed total angular momentum quantum number 
$s \in \{ 0, \tfrac{1}{2}, 1, \tfrac{3}{2}, 2, \dotsc \}$.
The usual arguments show that the eigenstates of $z$-angular momentum, $\ket{sm}$, form a ladder with $m = -s, -s+1, -s+2, \dotsc, s$.  The state with maximal $z$-angular momentum is $\ket{ss}$.  Since 
$\hat{S}_z \ket{ss} = s \ket{ss}$ and
$\hat{S}_x \ket{ss} = \hat{S}_y \ket{ss} = \ket{\mathrm{null}}$,
we see that $\ket{ss}$ is an eigenstate of the \emph{vector} spin operator in the sense that
	\begin{align}
	\hat{\SSS} \ket{ss} = s \eee_z \ket{ss}
	,
	\label{SpecificEigenrelation}
	\end{align}
where $\eee_z$ is the unit vector in the $z$ direction.  
Now, let $\sss$ be a vector of length $s$ and direction $(\theta,\phi)$, so that 
$\sss=(s,\theta,\phi)$ in spherical polars and 
$\sss=(s \sin \theta \cos \phi,s \sin \theta \sin \phi,s \cos \theta)$ in Cartesians.  
Define the \emph{spin coherent state} $\ket{\sss}$ as the state obtained by rotating $\ket{ss}$ counterclockwise by angle $\theta$ about the $y$ axis, and then by angle $\phi$ about the $z$ axis:
	\begin{align}
	\ket{\sss} 
	\equiv
	\ket{s \theta \phi} 
	&\equiv
	 e^{ i \phi \hat{S}_z } e^{ i \theta \hat{S}_y } \ket{ss}
	.
	\label{SCSDefinition}
	\end{align}
Using the Wigner $D$-matrix \cite{sakuraiBook}, we may write $\ket{\sss}$ explicitly as a linear combination of $\ket{sm}$ states:
	\begin{align}
	\ket{s \theta \phi} 
	&= \sum_{m=-s}^s 
		\sqrt{ \tfrac{(2s)!}{(s+m)!(s-m)!}  } 
		\big(  \cos \tfrac{\theta}{2} \big)^{s+m}
		\big(  \sin \tfrac{\theta}{2} \big)^{s-m}
		e^{-i m \phi} 
		 \ket{sm}
	.
	\label{WignerDMatrixFormula}
	\end{align}
	
From the properties of spin operators it can be shown that Eqs.~\eqref{SpecificEigenrelation} and \eqref{SCSDefinition} give
	\begin{align}
	\hat{\SSS} \ket{\sss} = \sss \ket{\sss}
	.
	\label{GeneralEigenrelation}
	\end{align}
In other words, the spin coherent state $\ket{\sss}$ is an eigenstate of the \emph{vector} spin operator with \emph{vector} eigenvalue $\sss$.  From here it is easy to show that
$
	\sss \cdot \hat{\SSS} \ket{\sss} = s^2 \ket{\sss}
$
and
$
	\bra{\sss} \hat{\SSS} \ket{\sss} = \sss
$,
and that the spin coherent states are normalized as
	\begin{align}
	\braket{\sss}{\sss} = 1.
	\end{align}

\section{Spin wavefunctions and orbital wavefunctions}
From Eq.~\eqref{WignerDMatrixFormula} we see that
	\begin{align}
	\braket{sm}{s \theta \phi} 
	&= 
		\sqrt{ \tfrac{(2s)!}{(s+m)!(s-m)!}  } 
		\big(  \cos \tfrac{\theta}{2} \big)^{s+m}
		\big(  \sin \tfrac{\theta}{2} \big)^{s-m}
		e^{-i m \phi} 
	.
	\end{align}
We define the ``spin wavefunction'' of a spin state $\ket{sm}$ as the ``coefficients'' of $\ket{sm}$ in the basis of spin coherent states $\ket{s \theta \phi}$, including a normalization factor:
	\begin{align}
	F_{sm} (\theta, \phi)
	&= \sqrt{ \tfrac{2s+1}{4\pi} } 
		\braket{s \theta \phi} {sm}
			\nonumber\\
	&= 
		\sqrt{ \tfrac{2s+1}{4\pi} \tfrac{(2s)!}{(s+m)!(s-m)!}  } 
		\big(  \cos \tfrac{\theta}{2} \big)^{s+m}
		\big(  \sin \tfrac{\theta}{2} \big)^{s-m}
		e^{i m \phi} 
	.
	\label{FsmFormula}
	\end{align}
For comparison, we also define ``orbital wavefunctions'' as the orbital angular momentum eigenfunctions for integer $l$ and $m$.  These are the well-known spherical harmonics\cite{sakuraiBook}, which can be written in terms of associated Legendre functions $P_l^m$:
	\begin{align}
	Y_{lm} (\theta, \phi)
	&= \sqrt{ \tfrac{2l+1}{4\pi} \tfrac{(l-m)!}{(l+m)!}  } 
		P_l^m (\cos\theta)
		e^{i m \phi} 
	.
	\label{YlmFormula}
	\end{align}
With these definitions, both types of wavefunctions are normalized such that
	\begin{align}
	\int_0^\pi d\theta~ \sin \theta
	\int_0^{2\pi} d\phi~
	\abs{ Y_{lm} (\theta, \phi) }^2
	&= 1
	,		\nonumber\\
	\int_0^\pi d\theta~ \sin \theta
	\int_0^{2\pi} d\phi~
	\abs{ F_{sm} (\theta, \phi) }^2
	&= 1
	.
	\end{align}
The spin wavefunction and orbital wavefunction have the same $\phi$ dependence, but the $\theta$ dependence is different.  The relationship between the two representations is vaguely reminiscent of the duality between position and momentum.  

\section{Visualizations}
We will borrow geographical techniques for visualizing functions over the surface of a sphere.  In particular, we will use the Hammer projection\cite{snyderBook}, which is an equal-area cartographic projection that maps the entire surface of the Earth (or any sphere) to the interior of an ellipse of semiaxes $\sqrt{8}$ and $\sqrt{2}$.  This may be thought of as making a cut along the ``International Dateline'' (the meridian $\phi=180^\circ$, so that the cut sphere is topologically equivalent to a flat sheet, and flattening the resulting shape into an ellipse.  The Hammer projection is described mathematically by the following transformations between $(\theta,\phi)$ and $(x,y)$:
	\begin{align}
		x = \frac{\sqrt{8} \sin \theta \sin \tfrac{\phi}{2}}{ 
				\sqrt{1 + \sin\theta \cos  \tfrac{\phi}{2}}
			}
	,\qquad
		y = \frac{\sqrt{2} \cos \theta }{
				\sqrt{1 + \sin\theta \cos  \tfrac{\phi}{2}}
			}
	,\qquad
		0 \leq \theta \leq \pi			
		\text{~and~}
		0 \leq \phi < 2\pi
	;
	\\
	\theta = \arccos \Big(  y\sqrt{1 - \tfrac{x^2}{16} - \tfrac{y^2}{4}		}   \Big)
	,\qquad
	\phi = 2\arctan \frac{ x\sqrt{1 - \tfrac{x^2}{16} - \tfrac{y^2}{4} 	}  }{
			4 (1 - \tfrac{x^2}{16} - \tfrac{y^2}{4}) - 2  
		}
	,\qquad
		\tfrac{x^2}{8} + \tfrac{y^2}{2} < 1
		.
	\label{HammerFormulas}
	\end{align}

	\begin{figure}
	  \begin{center}
			\includegraphics[width=0.85\columnwidth]{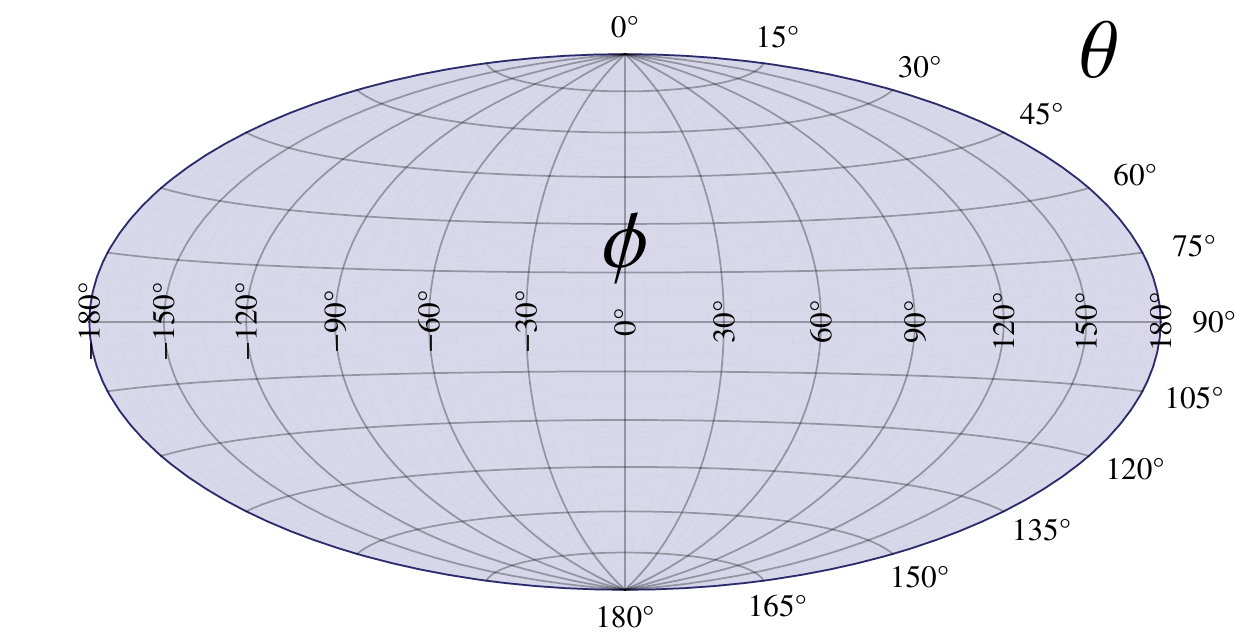}
	  \end{center}
	  \caption{
		Latitudes (lines of constant $\theta$)
		and meridians (lines of constant $\phi)$
		according to	the Hammer projection,
		which maps an entire spherical surface to a flat ellipse.
	  	\label{HammerFigure}
		}
	\end{figure}

	\begin{figure}
	\subfigure[$Y_{99}$]{
		\includegraphics[width=0.4\textwidth]{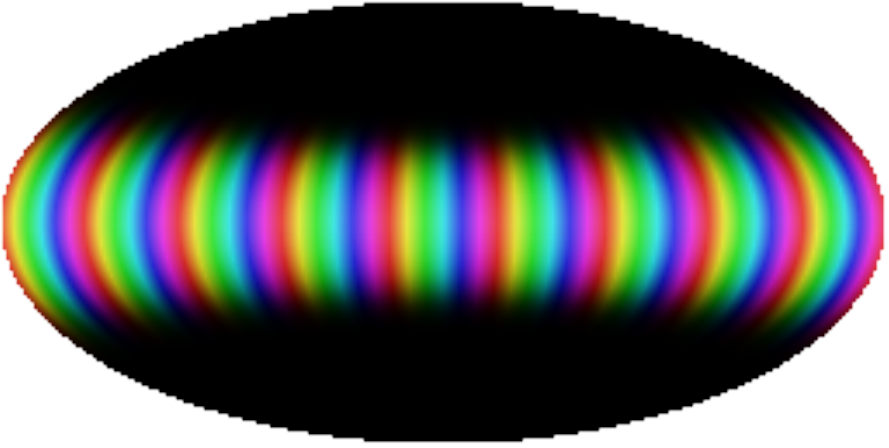}
		\label{Y99}
	}
	\subfigure[$F_{99}$]{
		\includegraphics[width=0.4\textwidth]{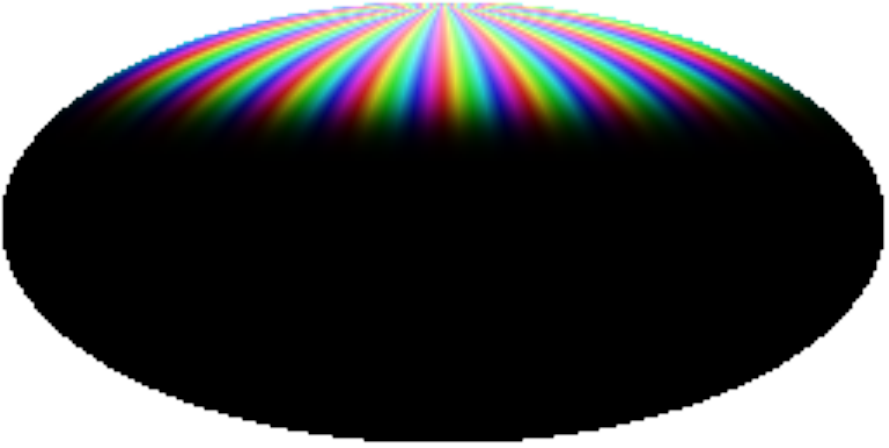}
		\label{F99}
	}
	\subfigure[$Y_{90}$]{
		\includegraphics[width=0.4\textwidth]{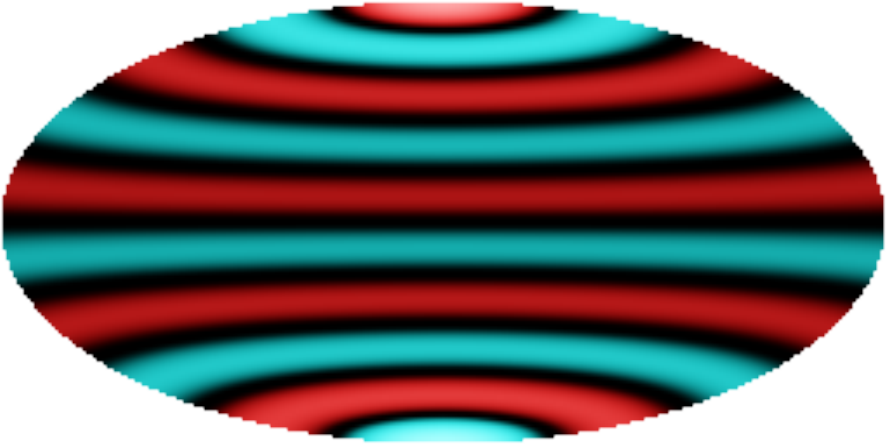}
		\label{Y90}
	} 
	\subfigure[$F_{90}$]{
		\includegraphics[width=0.4\textwidth]{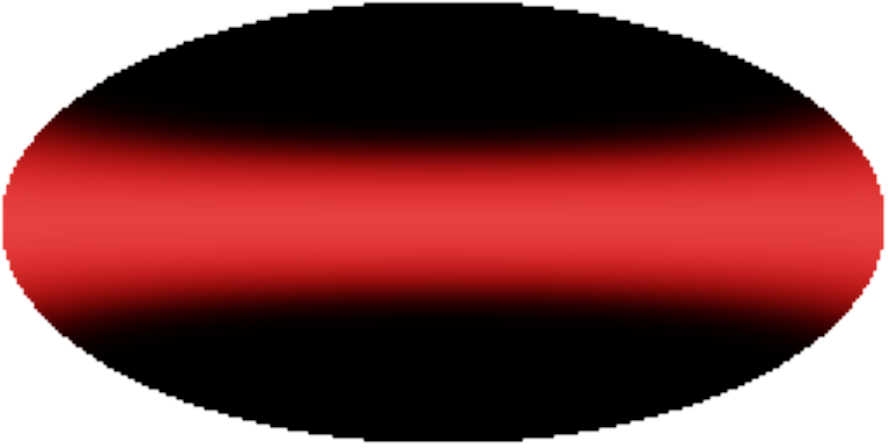}
		\label{F90}
	} 
	\subfigure[$Y_{96}$]{
		\includegraphics[width=0.4\textwidth]{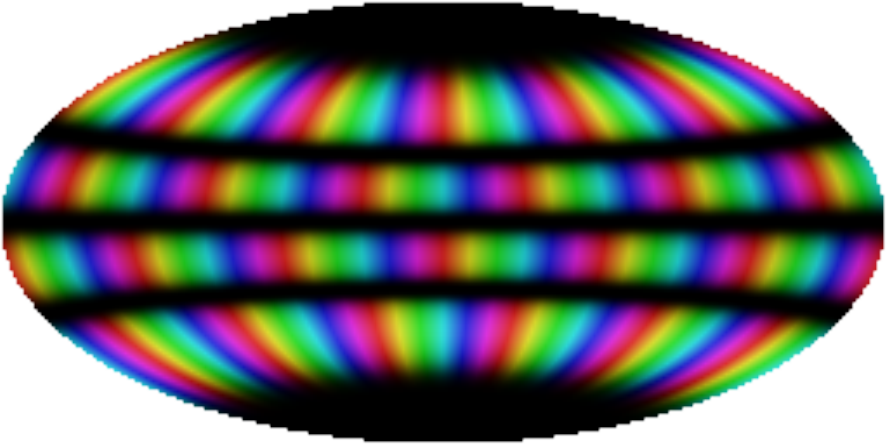}
		\label{Y96}
	} 
	\subfigure[$F_{96}$]{
		\includegraphics[width=0.4\textwidth]{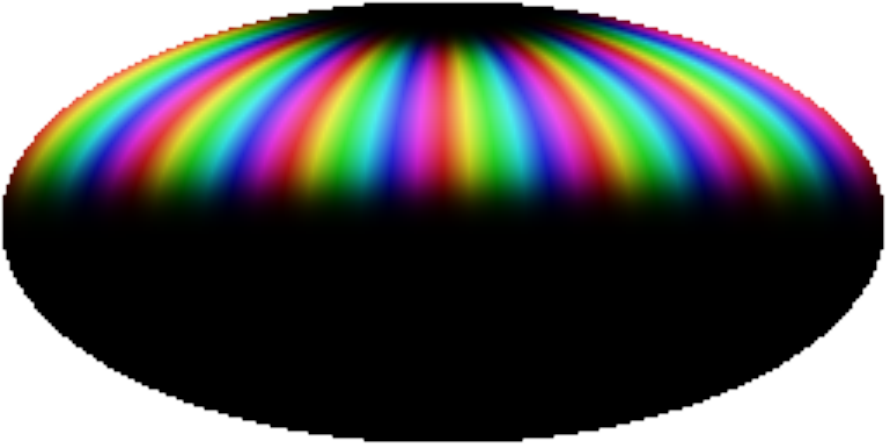}
		\label{F96}
	} 
	  \caption{
	  	Visualizations of ``orbital wavefunctions'' $Y_{lm} (\theta,\phi)$
				and ``spin wavefunctions'' $F_{sm} (\theta,\phi)$
				over the surface of the sphere
				according to the Hammer projection.
			Brightness indicates the magnitude of the complex wavefunction
				and hue indicates the argument.
	  	\label{YandF}
		}
	\end{figure}

Let us now compare spin and orbital wavefunctions for various states.
First consider the state of maximal $z$-angular momentum, $\ket{ss}$.  The orbital wavefunction is
	\begin{align}
	Y_{ss} (\theta, \phi)
	&= (-1)^s \sqrt{ \tfrac{(2s+1)!!}{ 2^s s!~ 4\pi } }  ~ \sin^s \theta ~ e^{i s \phi}
	\label{Yss}
	\end{align}
(Fig.~\ref{Y99}).  This represents travelling waves going around the equator of the sphere, which jives with the heuristic classical picture of a particle orbiting in a horizontal circle.  The orbital wavefunction is the probability amplitude for finding the particle at a position 
$\rrr=(r,\theta,\phi)$.  The spin wavefunction is
	\begin{align}
	F_{ss} (\theta, \phi)
	&= \sqrt{ \tfrac{2s+1}{ 4\pi } }  ~ \cos^{2s} \tfrac{\theta}{2} ~ e^{i s \phi}
	\label{Fss}
	\end{align}
(Fig.~\ref{F99}).  Since $\ket{ss}$ is identical to the spin coherent state 
$\ket{s \eee_z}$ with $\theta=\phi=0$, we might have expected that the spin wavefunction would be a Dirac delta function of the form $\delta(\theta) \delta(\phi)$.  However, because spin coherent states form an overcomplete non-orthonormal basis, the spin wavefunction is actually a smooth function with maximum amplitude near the north pole ($\theta=0$).  This can be interpreted in terms of a semiclassical spin vector that points toward the north pole on average, but undergoes quantum fluctuations away from this direction.  Larger values of $s$ lead to smaller quantum fluctuations.

Now consider the state $\ket{s0}$.  The orbital wavefunction is 
	\begin{align}
	Y_{s0} (\theta, \phi)
	&= \sqrt{ \tfrac{2s+1}{ 4\pi } } ~ P_s (\cos \theta)
	\label{Yss}
	\end{align}
(Fig.~\ref{Y90}), where $P_s$ is a Legendre polynomial.  This state has total angular momentum $s$, but its average $z$-angular momentum is zero.  Classically, this suggests that the angular momentum vector $\SSS$ lies in the $xy$ plane and that a particle executes circular orbits in a vertical plane perpendicular to $\SSS$.  Quantum mechanically, there are standing waves formed by the interference of northbound and southbound waves along every meridian.  

The spin wavefunction is
	\begin{align}
	F_{s0} (\theta, \phi)
	&= \sqrt{ \tfrac{2s+1}{ 4\pi }  \tfrac{(2s)!}{2^s s!}  } ~ \sin^{s} \theta	
	\label{Fss}
	\end{align}
(Fig.~\ref{F90}).  The plot can be understood as the distribution of a semiclassical spin whose quantum fluctuations allow it to explore the whole equator, as well as making excursions toward the ``tropics''.  This is an improvement over the textbook picture (Fig.~\ref{SpinCartoon}).  It is mathematically precise, and it captures extra nuances: not only does the spin precess in a circle due to quantum fluctuations, its ``latitude'' also fluctuates.

Finally, consider $\ket{96}$ as an example of a generic state.  For this state the orbital wavefunction (Fig.~\ref{Y96}) consists of travelling waves along several latitudes, whereas the spin wavefunction (Fig.~\ref{F96}) is concentrated near a single latitude.  This latitude corresponds to a vertical position $m$ on a sphere of radius $s(s+1)$, where $m=6$ and $s=9$.

\begin{figure}
	\subfigure[$Y_{lm}$]{
		\includegraphics[width=0.47\textwidth]{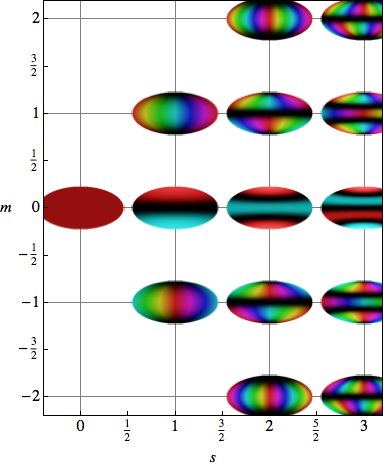}
		\label{Ylm}
	} 
	\subfigure[$F_{sm}$]{
		\includegraphics[width=0.47\textwidth]{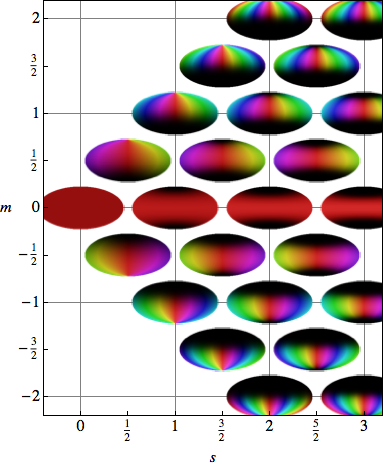}
		\label{Fsm}
	} 
	  \caption{
	  	Tables of visualizations 
			of ``orbital wavefunctions'' $Y_{lm} (\theta,\phi)$
			and ``spin wavefunctions'' $F_{sm} (\theta,\phi)$.
			The latter are defined for both integer and half-integer $s$.
	  	\label{YandFTable}
		}
	\end{figure}

The orbital wavefunction $Y_{lm}$ is only meaningful when $l$ and $m$ are integers.  If $l$ is a half-integer, $Y_{lm}$ diverges at the poles and is generallly non-normalizable, due to the Legendre functions in Eq.~\eqref{YlmFormula}.  In contrast, the spin wavefunction $F_{sm}$ is well-defined even for half-integer values of $s$ and $m$, as can be seen from	Eq.~\eqref{FsmFormula} and Fig.~\ref{YandFTable}.  This is the key advantage of spin wavefunctions.

A careful reader will notice that if $s$ is a half-integer, the function $F_{sm} (\theta, \phi)$ is discontinuous at the ``International Dateline'' $\phi=\pi$: upon crossing this branch cut, the spin wavefunction changes by a factor of $-1$.  This is not a bug, but a feature!  It illustrates the peculiar nature of spinor rotation: rotation by $2\pi$ gives a factor of $-1$, and a spinor is only invariant under a full $4\pi$ rotation.

\section{Spin wavefunction of a spin coherent state}
\begin{figure}
		\subfigure[$F_{60^\circ,-45^\circ}$]{
			\includegraphics[width=0.3\textwidth]{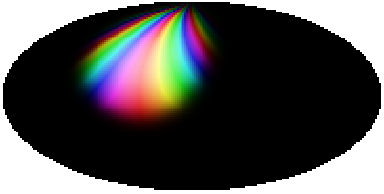}
				\label{Fscs1}
		} 
		\subfigure[$F_{60^\circ,45^\circ}$]{
			\includegraphics[width=0.3\textwidth]{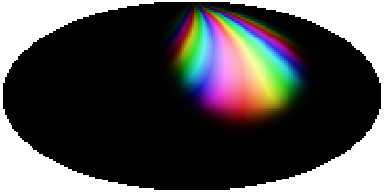}
				\label{Fscs2}
		} 
		\subfigure[$F_{60^\circ,135^\circ}$]{
			\includegraphics[width=0.3\textwidth]{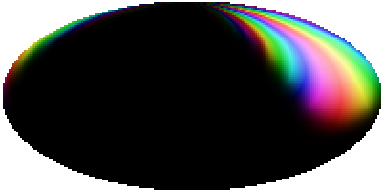}
				\label{Fscs3}
		} 
	  \caption{
		  Spin wavefunctions $F_{\theta'\phi'} (\theta,\phi)$
				for three spin coherent states $\ket{s\theta'\phi'}$
			with spin quantum number is $s=19/2$,
				colatitude parameter $\theta'=60^\circ$,
				and azimuthal parameters $\phi'=-45^\circ, 45^\circ, 135^\circ$.
		  \label{Fscs}
		}
	\end{figure}

So far we have considered ``spin wavefunctions'' for spin angular momentum eigenstates $\ket{sm}$.  Now let us consider the spin wavefunction for a spin coherent state $\ket{\sss'} = \ket{s \theta' \phi'}$:
	\begin{align}
	F_{\theta'\phi'} (\theta, \phi)
	&\equiv 
	\sqrt{ \tfrac{2s+1}{ 4\pi } }
	\braket{s\theta\phi}{s \theta' \phi'}
			\nonumber\\
	&= 
		\sqrt{ \tfrac{2s+1}{ 4\pi } }
		\sum_{m=-s}^s
		\tfrac{(2s)!}{(s+m)!(s-m)!}  
		\big(  \cos \tfrac{\theta}{2} \cos \tfrac{\theta'}{2} \big)^{s+m}
		\big(  \sin \tfrac{\theta}{2} \sin \tfrac{\theta'}{2} \big)^{s-m}
		e^{i m (\phi - \phi')}  
	.
	\label{Fthetaphithetaphi}
	\end{align}
Although the form of Eq.~\eqref{Fthetaphithetaphi} is not illuminating, the plots in Fig.~\ref{Fscs} show that the ``spin wavefunction'' $	F_{\theta'\phi'} (\theta, \phi)$ has largest magnitude near $\theta=\theta'$ and $\phi=\phi'$, as one would expect.  The ``spread'' in the wavefunction is inversely proportional to $s$, so that the limit $s\rightarrow\infty$ is indeed the semiclassical limit.

\section{Time-dependent spin wavefunctions}
A spin in a constant magnetic field $\BBB$ obeys the Hamiltonian $\hat{H} = -\gamma \BBB \cdot \hat{\SSS}$ (where $\gamma$ is a gyromagnetic ratio)\cite{griffithsBook}.  Ehrenfest's theorem shows that the average spin precesses around the direction of $\BBB$ at the Larmor frequency $f_L = \gamma B/2\pi$.  
Furthermore, it can be shown that if the initial state is a spin coherent state, $\ket{\sss}$, then the state at a later time $t$ is also a spin coherent state with a rotated vector $\sss(t)$ as well as a phase factor.  This implies that Larmor precession can be visualized in the classroom by animating the time-dependent spin wavefunction $F(\theta,\phi,t)$.  Successive frames in such an animation might look like Fig.~\ref{Fscs}.

\section{Further discussion}
It is well known that the orbital wavefunctions for $\ket{lm}$ states, which are the spherical harmonics $Y_{lm} (\theta, \phi)$, are polynomials in $x$, $y$, and $z$, where $x=\sin\theta\cos\phi$, $y=\sin\theta\sin\phi$, and $z=\cos\theta$.  Starting from Eq.~\ref{FsmFormula}, it can be shown that the spin wavefunctions for $\ket{sm}$ states, $F_{sm} (\theta, \phi)$, are square roots of rational functions in $x$, $y$, and $z$:
	\begin{align}
	F_{sm} (\theta, \phi)
	&= 
		\sqrt{ \tfrac{2s+1}{4\pi} \tfrac{(2s)!}{(s+m)!(s-m)! 4^s}  } 
		(1+z)^{s/2}
		(1-z)^{s/2-m}
		(x+iy)^m
	.
	\end{align}
The spin wavefunctions for coherent states $\ket{s\theta\phi}$ can be expressed in a similar form:
	\begin{align}
	F_{\theta'\phi'} (\theta, \phi)
	&= 
		\tfrac{2s+1}{4^s 4\pi} 
		\sum_{m=-s}^s
		\tfrac{(2s)!}{(s+m)!(s-m)!}
		(1+z)^{s/2}
		(1+z')^{s/2}
		(1-z)^{s/2-m}
		(1-z')^{s/2-m}
		(x+iy)^m
		(x'-iy')^{m}
	,
	\end{align}
where $x'$, $y'$, and $z'$ correspond to $\theta'$ and $\phi'$.  The sum can be written in closed form in terms of hypergeometric functions, but this is not illuminating.

The spin coherent states form a massively overcomplete non-orthonormal basis.  Thus, the spin wavefunction $F(\theta,\phi)$ contains a large amount of redundant information.
However, we conjecture that the $F_{sm} (\theta, \phi)$ for \emph{all} $s=0, \tfrac{1}{2}, 1, \tfrac{3}{2}, 2, \dotsc$ and $m=-s,-s+1,\dotsc,s$ may form a complete, non-redundant basis for functions on the sphere, just like the $Y_{lm} (\theta, \phi)$.  If this is true, it would allow an \emph{arbitrary} wavefunction to be expanded as $F(\theta,\phi) = \sum_{sm} c_{sm} F_{sm} (\theta,\phi)$.  We are not aware whether this has been proven.  
At the time of writing it is also unclear whether the time-dependent Schr{\"o}dinger equation for $F(\theta,\phi;t)$ can be written down in differential form, or if it is inherently an integrodifferential equation.  Measurement-induced collapse of the ``spin wavefunction'' can certainly be discussed within our picture, although this may not serve a useful pedagogical purpose.

\section{Closing remarks}
We have developed the concept of the ``spin wavefunction'' for spin-$s$ spins, using the basis of spin coherent states.  This works for both integer and half-integer values of $s$.  We provide explicit formulas and striking visualizations of spin eigenstates $\ket{sm}$, spin coherent states $\ket{\sss}$, and Larmor precession.
We also demonstrate that cartographic projections such as the Hammer projection are useful for visualizing wavefunctions defined on spherical surfaces.

Students bring a variety of learning styles to the classroom \cite{montgomery1999}.
Some take well to a deductive approach going from general theorems to specific phenomena, 
whereas others prefer an inductive approach starting with concrete examples.
We feel that the spin wavefunction visualizations presented here will be very useful for reaching out to the latter class of students.





\end{document}